\documentstyle[preprint,aps,epsfig]{revtex}
\tightenlines
\begin{document}
\draft
%%%%%%%%%%%%%%%%%%%%%%%%%%%%%%%%%%%%%%%%%%%%%%%%%%%%%%%%%%%%%%%%%%%%%
%%%%%%%%%%%%%%%%%%%%%         Title       %%%%%%%%%%%%%%%%%%%%%%%%%%%
%%%%%%%%%%%%%%%%%%%%%%%%%%%%%%%%%%%%%%%%%%%%%%%%%%%%%%%%%%%%%%%%%%%%%

\title{\begin{flushright}
          {\small IFT-P.044/99 \,\, gr-qc/9905096}
       \end{flushright}
       Search for semiclassical--gravity effects in relativistic stars}

%%%%%%%%%%%%%%%%%%%%%%%%%%%%%%%%%%%%%%%%%%%%%%%%%%%%%%%%%%%%%%%%%%%%%       
%%%%%%%%%%%%%%%%%%%%     Authors & Addresses  %%%%%%%%%%%%%%%%%%%%%%%
%%%%%%%%%%%%%%%%%%%%%%%%%%%%%%%%%%%%%%%%%%%%%%%%%%%%%%%%%%%%%%%%%%%%%

\author{Daniel A.T. Vanzella\footnote{vanzella@ift.unesp.br}
        and George E.A. Matsas\footnote{matsas@ift.unesp.br}}
\address{Instituto de F\'\i sica Te\'orica, Universidade Estadual Paulista,\\
         Rua Pamplona 145, 01405-900, S\~ao Paulo, SP\\
         Brazil}
\def\baselinestretch{1.5}
\maketitle

%%%%%%%%%%%%%%%%%%%%%%%%%%%%%%%%%%%%%%%%%%%%%%%%%%%%%%%%%%%%%%%%%%%%%%%
%%%%%%%%%%%%%%%%%%%%%           Abstract         %%%%%%%%%%%%%%%%%%%%%%
%%%%%%%%%%%%%%%%%%%%%%%%%%%%%%%%%%%%%%%%%%%%%%%%%%%%%%%%%%%%%%%%%%%%%%%

\begin{abstract}

We discuss the possible influence of gravity in the neutronization process, 
$\; p^+ \;e^- \to n \; \nu_e$, which is particularly important as a cooling
mechanism of neutron stars.  Our approach is semiclassical in the sense that
leptonic fields are quantized on a classical background spacetime, while
neutrons and protons are treated as excited and unexcited nucleon states,
respectively.  We expect gravity to have some influence wherever the energy
content carried by the in--state is barely above the neutron mass.  
In this case
the emitted neutrinos would be soft enough to have a wavelength of the
same order as the space curvature radius.
\end{abstract}

\newpage

%%%%%%%%%%%%%%%%%%%%%%%%%%%%%%%%%%%%%%%%%%%%%%%%%%%%%%%%%%%
%!!!!!!!!!!!!!!!!!!!!     paper    !!!!!!!!!!!!!!!!!!!!!!!%
%%%%%%%%%%%%%%%%%%%%%%%%%%%%%%%%%%%%%%%%%%%%%%%%%%%%%%%%%%%

The inner structure of neutron stars has attracted much attention
of relativists and particle and nuclear physicists 
since there still remain many subtle points to be better understood
(see, e.g., \cite{NS} and references therein). 
It would be interesting, thus, to investigate how gravity could influence 
quantum  processes  which occur in their interior. The gravitational 
field may be of some significance to quantum phenomena wherever
they involve particles with wavelengths of the same order as the space
curvature radius. Processes dealing with soft particles 
are very promising since their wavelengths 
may be arbitrarily large. Here we focus on the neutronization process,      
$p^+ \;e^- \to n \; \nu_e$, which is an important cooling mechanism for 
neutron stars with temperatures up to about $10^9$~K. 
Our approach is essentially semiclassical in the sense that leptonic 
fields are quantized on a classical background spacetime, while 
neutrons and protons are treated as excited and unexcited nucleon 
states, respectively. We will use natural  units $G = \hbar = c = k_{B} = 1$ 
throughout this paper.

Field quantization in the Schwarzschild spacetime 
is not easy to accomplish~\cite{JC}. We shall simplify 
the problem by simulating the  Schwarzschild 
spacetime by a two--dimensional noninertial frame described by 
the Rindler wedge. The Rindler wedge is a static spacetime
defined by the line element
\begin{equation}
ds^2 = a^2 u^2 d\tau^2 - du^2 \;  
\label{LE}
\end{equation}
with $0<u<+\infty$ and $-\infty<\tau<+\infty$, where $a$ 
``characterizes'' the frame acceleration, i.e., 
$a \equiv \sqrt{a_\mu a^\mu} = {\rm const}$ is the
proper acceleration of the worldline which has
$\tau$ as its proper time, namely, $u=a^{-1}$.

We would like to consider the case in which the nucleons lie 
approximately static during the reaction at some fixed point in the 
star. In principle, this poses no problem since the whole process 
takes place in the presence of a medium and not in the vacuum.
The location where the reaction happens will be specified by the 
nucleon proper acceleration. 
Thus, in our simplified 
model the reacting nucleons will be described by a uniformly
accelerated current in the Rindler wedge with constant proper 
acceleration $a$: 
\begin{equation}
j^\mu = q u^\mu \delta (u-a^{-1}) ,
\label{C}
\end{equation}
where $q$ is a small coupling constant
and $u^\mu = (a,0)$ is the nucleon 
four--velocity. Next, in order to allow the
current above to describe the proton--neutron transition, we shall
consider the nucleon  as a two--level system~\cite{U}. 
In this vein, neutrons $|n \rangle$ and
protons $|p  \rangle$ will be excited and unexcited 
eigenstates of the nucleon  Hamiltonian $\hat H$: 
\begin{equation}
\hat H |n \rangle = m_n |n \rangle \;\; ,\;\;
\hat H |p \rangle = m_{p } |p  \rangle \;\; ,
\end{equation}
where $m_n$ and $m_{p }$ are  the neutron and proton masses,
respectively. Hence current (\ref{C}) will be replaced by 
\begin{equation}
\hat j^\mu = \hat q(\tau) u^\mu \delta (u-a^{-1}) \;\;,
\label{CI}
\end{equation}
where 
$ \;\hat q(\tau )\equiv 
\exp({i\hat H \tau}) \; \hat q_0\; \exp({-i\hat H \tau}) $
is a Hermitian monopole. 
The two--dimensional  Fermi constant 
$G_F \equiv |\langle p | \hat q_0 | n \rangle | = 9.918 \times 10^{-13}$ 
is determined~\cite{MV} by imposing that the  mean proper 
lifetime of inertial neutrons is $887$ s~\cite{PDG}.

In order to calculate the neutronization rate we shall quantize the 
leptonic  field in the Rindler wedge. 
The leptonic field is expressed as \cite{SMG} 
\begin{equation}
\hat \Psi(\tau,u)= \sum_{\sigma = \pm } \int_{0}^{+\infty} d\omega
\left( \hat b_{\omega \sigma} \psi_{\omega \sigma}(\tau,u)
       + \hat d^\dagger_{\omega \sigma} \psi_{-\omega -\sigma}(\tau,u) 
\right) ,
\label{PSA}
\end{equation}
where 
$\psi_{\omega \sigma}(\tau,u)=f_{\omega \sigma}(u) e^{-i \omega \tau}$
are positive ($\omega >0$) and negative ($\omega <0$)
frequency solutions of the Dirac equation,
with respect to the boost Killing field 
$\partial / \partial \tau$, with polarizations $\sigma = \pm$. 
We recall that Rindler frequencies may assume arbitrary positive real values.
In particular there are massive Rindler particles with arbitrarily small
frequencies. (See Ref. \cite{HMS} for a discussion on zero-frequency 
Rindler particles.)
Here
\begin{eqnarray}
f_{\omega +} (u) &=&
A_+ 
\left(
\begin{array}{c}
K_{i \omega/a + 1/2}(mu) +
i K_{i \omega/a - 1/2}(mu)\\
0\\
-K_{i \omega/a + 1/2}(mu) +
i K_{i \omega/a - 1/2}(mu)\\
0
\end{array}
\right) \; ,
\\
f_{\omega -} (u)
&=&
A_- 
\left(
\begin{array}{c}
0\\
K_{i \omega/a + 1/2}(mu) +
i K_{i \omega/a - 1/2}(mu)\\
0\\
 K_{i \omega/a + 1/2}(mu) -
i K_{i \omega/a - 1/2}(mu)
\end{array}
\right) \; ,
\end{eqnarray}
where $m$ is the lepton mass and the normalization constants 
\begin{equation}
A_+=A_-=\left[\frac{m \cosh (\pi \omega/a)}{2\pi^2 a}\right]^{1/2}\; 
\label{CONST}
\end{equation} 
were chosen such that the annihilation and creation operators  satisfy
the following simple anticommutation relations
\begin{equation}
\{\hat b_{\omega \sigma},\hat b^\dagger_{\omega' \sigma'}\}=
\{\hat d_{\omega \sigma},\hat d^\dagger_{\omega' \sigma'}\}=
\delta(\omega-\omega') \; \delta_{\sigma \sigma'} \; ,
\label{ACR2}
\end{equation}
\begin{equation}
\{\hat b_{\omega \sigma},\hat b_{\omega' \sigma'}\}=
\{\hat d_{\omega \sigma},\hat d_{\omega' \sigma'}\}=
\{\hat b_{\omega \sigma},\hat d_{\omega' \sigma'}\}=
\{\hat b_{\omega \sigma},\hat d^\dagger_{\omega' \sigma'}\}=
0 \;\; .
\end{equation} 

Now we are ready to calculate the neutronization 
amplitude 
\begin{equation}
{\cal A}  = 
\; \langle  n \vert \otimes \langle \nu_{\omega_\nu \sigma_\nu} \vert \;
\hat S_I \;
\vert e^-_{\omega_{e} \sigma_{e}} \rangle \otimes \vert p  \rangle\; ,
\label{ACA}
\end{equation}
where we minimally couple 
the nucleon current (\ref{CI}) to the leptonic fields $\hat \Psi_e$
and $\hat \Psi_\nu$ through the Fermi interaction action 
\begin{equation}
\hat S_I = \int d^2x \sqrt{-g}\; \hat j_\mu 
           (\hat{\bar \Psi}_\nu \gamma^\mu_R \hat \Psi_e +
            \hat{\bar \Psi}_e \gamma^\mu_R \hat \Psi_\nu ) \; .
\label{S}
\end{equation} 
In the Rindler wedge 
$\gamma_R^\mu \equiv (e_\alpha)^\mu \gamma^\alpha$ with tetrads 
$(e_0)^\mu =u^{-1} \delta_0^\mu$ and  
$(e_i)^\mu = \delta_i^\mu$, where $\gamma^\alpha$
are the usual Dirac matrices.
By using Eq.~(\ref{S}) in Eq.~(\ref{ACA}),
we obtain the following amplitude:
\begin{equation}
{\cal A}_{ac} = 
G_F \int_{-\infty}^{+\infty} d\tau \;e^{i \Delta m \tau} 
\langle \nu_{\omega_\nu \sigma_\nu} \vert
\hat\Psi^\dagger_\nu  (\tau,a^{-1}) \hat \Psi_e (\tau,a^{-1})
\vert e^-_{\omega_{e} \sigma_{e}} \rangle\; ,
\label{V}
\end{equation}
where $\Delta m \equiv m_n - m_p$. Next, by using Eq.~(\ref{PSA}), we obtain
\begin{equation}
{\cal A}_{ac} = G_F \;\delta_{\sigma_{e},\sigma_\nu}
\int_{-\infty}^{+\infty} d\tau\; e^{i \Delta m \tau}
\psi^\dagger_{\omega_\nu\sigma_\nu} (\tau,a^{-1}) 
\;\psi_{\omega_{e}\sigma_{e}}(\tau,a^{-1})\;.
\label{V1}
\end{equation}
Using now explicitly $\psi_{\omega \sigma } (\tau,u)$ to perform the integral, 
we obtain
\begin{eqnarray}
{\cal A}_{ac} & = & \frac{4G_F}{\pi a}\sqrt{m_em_\nu \cosh (\pi
\omega_{e}/a)\cosh (\pi \omega_\nu/a)} \nonumber \\
& \times & 
Re\left[ K_{i\omega_\nu/a-1/2}(m_\nu /a)\; K_{i\omega_{e}/a +1/2} (m_e /a) 
\right] 
\delta_{\sigma_{e}, \sigma_\nu} \delta( \omega_{e}-\omega_\nu-\Delta m)\;.
\label{V2}
\end{eqnarray}

This result will be used to calculate the total reaction rate    
\begin{equation}
\Gamma_{ac} (a)
\equiv
\frac{1}{\tilde{\tau}}
\sum_{\sigma_{e}=\pm} \sum_{\sigma_{\nu}=\pm}
\int_0^{+ \infty} d \omega_e \int_0^{+ \infty} d \omega_\nu
|{\cal A}_{ac}|^2
n_F(\omega_{e}, T_{e}) [1-n_F(\omega_\nu, T_{\nu})]\;,
\label{AP1}
\end{equation}
where $\tilde{\tau} = 2\pi \delta(0)$ is the total nucleon 
proper time~\cite{IZ},
and $ n_F(\omega, T) \equiv 1/ [1+exp(\omega /T)] $ is the usual
fermionic thermal factor. We shall consider further two  
cases. In the first one, we assume $T_{e}= 10^9$ K and
$T_\nu = 0$ K, i.e., the neutron star would be cold enough to be 
transparent to the neutrinos. In the second one, we assume 
$T_{e}= T_\nu = 10^{10}$ K, i.e., electrons and neutrinos would be 
in thermal equilibrium. By using Eq.~(\ref{V2}) in Eq.~(\ref{AP1}), we obtain 
\begin{eqnarray}
\Gamma_{ac} (a)
 =   
\frac{4G_F^2 m_e m_\nu }{\pi^3 a^2 }  
\int_{\Delta m}^{+\infty} & d\omega_{e}&
\frac{\cosh [\pi \omega_e/a] 
      \cosh [\pi (\omega_e - \Delta m)/a] 
      \exp[(\omega_e - \Delta m)/2 T_\nu]}
     {\cosh [\omega_e/2 T_e] 
      \cosh [(\omega_e - \Delta m)/2 T_\nu ]
      \exp[\omega_e/2 T_e]}
\nonumber \\
&\times &
\left\{Re\left[ K_{i(\omega_{e}-\Delta m)/a -1/2} (m_\nu/a)
                K_{i\omega_{e}+1/2}(m_e/a)\right] \right\}^2 
                                \; .
\label{GAMMA}
\end{eqnarray}
As a final step, we take the limit $m_\nu \to 0$ in Eq.~(\ref{GAMMA}) 
(see Ref.~\cite{GR}):
\begin{eqnarray}
\Gamma_{ac} (a)
= 
\frac{G_F^2 m_e }{\pi^2 a }
\int_{\Delta m}^{+\infty} 
& d\omega_{e} &
\frac{ \cosh [\pi \omega_e/a] \exp[(\omega_{e} -\Delta m)/2 T_\nu]}
{\cosh [\omega_e/2 T_e] \cosh[(\omega_{e} -\Delta m)/2T_\nu ] 
\exp[\omega_e/2 T_e]}
\nonumber \\
&\times &
K_{i\omega_{e}/a + 1/2} (m_e/a)K_{i\omega_{e}/a - 1/2} (m_e/a)\; .
\label{R5}
\end{eqnarray}

In order to compare the reaction rate above with the usual one obtained 
in inertial frames, we calculate  next the reaction rate for $a = 0$
using plain quantum field theory in Minkowski spacetime.
This will  be used also as a consistency check since we will  
compare it with the $a \to 0$ limit obtained from Eq.~(\ref{R5}). 

Let us briefly outline the Minkowski 
calculation.  The leptonic fields will be expressed 
in terms of the usual Minkowski coordinates
$(t,z)$ as   
\begin{equation}
\hat \Psi(t,z)= \sum_{\sigma = \pm } \int_{-\infty}^{+\infty} dk
\left( \hat b_{k \sigma} \psi^{(+\omega)}_{k \sigma} (t,z)
     + \hat d^\dagger_{k \sigma} \psi^{(-\omega)}_{-k -\sigma} (t,z) 
\right)\;,
\label{FF}
\end{equation}
where $ \hat b_{k \sigma} $ and $ \hat d^\dagger_{k \sigma} $ 
are annihilation and creation operators of fermions
and antifermions, respectively, with momentum $k$ and
polarization $\sigma$. In the inertial frame, energy,
momentum and mass $m$ are related as usual: 
$\omega=\sqrt{k^2+m^2}>0$. 
$ \psi^{(+\omega)}_{k \sigma} (t,z)$ 
and $ \psi^{(-\omega)}_{k \sigma} (t,z)$
are positive and negative frequency solutions of the Dirac equation
with respect to $\partial/{\partial t}$, respectively.
In the Dirac representation (see, e.g., Ref.~\cite{IZ}), we find
\begin{equation}
\psi^{(\pm \omega)}_{k +} (t,z) =
 \frac{e^{i(\mp \omega t + kz)}}{\sqrt{2\pi}}
\left(
\begin{array}{c}
\pm \sqrt{(\omega \pm m)/2\omega} \\
0\\
k/\sqrt{2\omega(\omega \pm m)}\\
0
\end{array}
\right) \;\; 
\label{NM1}
\end{equation}
and
\begin{equation}
\psi^{(\pm \omega)}_{k -} (t,z) = 
\frac{e^{i(\mp \omega t + kz)}}{\sqrt{2\pi}}
\left(
\begin{array}{c}
0\\
\pm \sqrt{(\omega \pm m)/2\omega} \\
0\\
-k/\sqrt{2\omega(\omega \pm m)}
\end{array}
\right) \;\; ,
\label{NM2}
\end{equation}
where the normalization constants were chosen such that the
creation and annihilation operators satisfy
\begin{equation}
\{\hat b_{k \sigma},\hat b^\dagger_{k' \sigma'}\}=
\{\hat d_{k \sigma},\hat d^\dagger_{k' \sigma'}\}=
\delta(k-k') \; \delta_{\sigma \sigma'} 
\label{ACR}
\end{equation}
and
\begin{equation}
\{\hat b_{k \sigma},\hat b_{k' \sigma'}\}=
\{\hat d_{k \sigma},\hat d_{k' \sigma'}\}=
\{\hat b_{k \sigma},\hat d_{k' \sigma'}\}=
\{\hat b_{k \sigma},\hat d^\dagger_{k' \sigma'}\}=
0 \;\; .
\end{equation}

The neutronization amplitude for inertial nucleons
in the Minkowski spacetime,
\begin{equation}
{\cal A}_{in} = G_F
\int_{-\infty}^{+\infty} dt\; e^{i \Delta m t}
{\psi^{(+\omega_\nu)}_{k_\nu \sigma_\nu}}^\dagger (t,0) 
\;\psi^{(+\omega_e)}_{k_e \sigma_{e}}(t,0) \; ,
\label{V3}
\end{equation}
is calculated by using the interaction action (\ref{S}) 
in Eq.~(\ref{ACA}), where $\gamma^\mu_R$ is
replaced by the usual  $\gamma^\mu$ Dirac matrices,
and the current is given by
$\hat j^\mu = \hat q(t) v^\mu \delta (z)$
with $v^\mu =(1,0)$.
This leads us straightforwardly to the following 
neutronization rate for inertial nucleons: 
\begin{equation}
{\Gamma}_{in}
=
\frac{2 G_F^2}{\pi}
\int_{L}^{+ \infty} d k_e
\frac{e^{(\omega_e - \Delta m)/T_\nu}}{(1+e^{{\omega}_e/T_e} )
[1+e^{({\omega}_e-\Delta m )/T_\nu}]}
 \; ,
\label{DP6}
\end{equation}
where $m_\nu = 0$, $L \equiv \sqrt{\Delta m^2 - m_e^2} $,  and we
recall that $\omega_e \equiv \sqrt{k_e^2 + m_e^2}$. 

In order to clearly analyze the influence of the
frame acceleration on the
neutronization process, let us use 
Eqs.~(\ref{R5}) and (\ref{DP6}) to define the following
relative reaction rate:
\begin{equation}
{\cal R} (a) \equiv \frac{{\Gamma}_{ac} (a) - 
{\Gamma}_{in}}{{\Gamma}_{in}} \; .
\label{R}
\end{equation}
In Figs.~1 and 2 we plot ${\cal R} (a) $ for the two aforementioned 
cases: (i) $\;T_e=10^{9}$ K and $T_{\nu}=0$ K,
and (ii) $\;T_e=T_{\nu}=10^{10}$ K. Firstly we  note
from the figures that ${\Gamma}_{ac}\; (a \to 0)$ is
in agreement with the expression obtained for
${\Gamma}_{in}$ since ${\cal R} (a \to 0) \to 0$.
Figs.~1 and 2 exhibit a complicated oscillatory pattern
up to $a \approx 1$ MeV. Indeed, the frame
acceleration plays its most important role in this region:
$|{\cal R}(a)|$ reaches about 30\% and 10\%
for cases (i) and (ii), respectively.
For large enough accelerations, $a \gg \Delta m,\;T_e$,
we obtain from Eq.~(\ref{R5})
an asymptotic expression for  ${\Gamma}_{ac}$, namely,
\begin{equation}
{\Gamma}_{ac} (a \gg \Delta m,\;T_e) \approx 
\frac{2 G_F^2}{\pi}
\int_{\Delta m}^{+ \infty} d {\omega}_e
\frac{e^{(\omega_e - \Delta m)/T_\nu}}{(1+e^{{\omega}_e/T_e} )
[1+e^{({\omega}_e-\Delta m )/T_\nu}]} \; .
\label{GINFTY}
\end{equation}
By using Eq.~(\ref{GINFTY}) in Eq.~(\ref{R}), we can compute
the asymptotic
relative reaction rate, namely,  ${\cal R} (a \gg \Delta m,\;T_e)$. 
We find that ${\cal R} (a \gg \Delta m,\;T_e) \approx -7.2 \%$ and
${\cal R} (a \gg \Delta m,\;T_e) \approx -3.5 \%$ 
for  cases (i) and (ii), respectively, i.e., according to our toy model,
ultrahigh accelerations damp the neutronization rate by a few percents.

In summary, we have looked for gravity effects in the neutronization
process which frequently occurs in the interior of neutron stars.
The reaction rate obtained by means of a simplified model 
exhibits a complicated oscillatory pattern up to
$a \approx 1$ MeV. Afterwards it tends to an asymptotic value
which indicates that the reaction is somewhat damped.  
We note that proper accelerations of the order $a \approx 1$ MeV
are much beyond what would  be expected in the interior of relativistic stars.
Just for sake of comparison, protons at LHC/CERN will be under accelerations
of about $10^{-8}$~MeV. We emphasize, however, that only a 
four--dimensional Schwarzschild 
calculation would be realistic enough to precisely determine the whole 
influence of gravity in the neutronization reaction and other similar 
processes.  In a more realistic calculation, for instance, 
effects due to the {\em space curvature} itself, 
which is absent here, should show up
wherever the emitted neutrinos are soft 
enough to ``feel'' the global background geometry. 
In this case, even reactions taking place at the
star core, where $a \approx 0$, would be influenced
by gravity. More detailed investigations on the  role played
by gravity in particle processes occuring in relativistic
stars would be welcome.

\begin{flushleft}
{\bf{\large Acknowledgments}}
\end{flushleft}
D.V. was fully supported by
Funda\c c\~ao de Amparo \`a Pesquisa do Estado de S\~ao Paulo
while G.M. was  partially supported by
Conselho Nacional de Desenvolvimento Cient\'\i fico e 
Tecnol\'ogico.

\newpage

\begin{center}
Figure Captions
\end{center}
\vskip 1 truecm

FIG. 1: The relative reaction rate ${\cal R} (a)$ is plotted
as a function of the frame acceleration $a$ for 
temperatures
$T_e=10^9$ K and $T_\nu=0$ K. Note that ${\cal R} 
(a \to 0)\to 0$, as expected. After an oscillatory regime
the relative reaction rate tends to the asymptotic value 
${\cal R} (a \gg \Delta m,\;T_e) \approx -7.2\%$.
The maximum value reached by $|{\cal R} (a)|$ is
about 30\%.

\vskip 1 truecm

FIG. 2: The relative reaction rate ${\cal R} (a)$ is plotted
as a function of the frame acceleration $a$ for 
temperatures
$T_e=T_\nu=10^{10}$ K.
After an oscillatory regime the 
relative reaction rate tends to the asymptotic value 
${\cal R} (a \gg \Delta m,\;T_e) \approx -3.5\%$.
The maximum value reached by $|{\cal R} (a)|$
is about 10\%.

\newpage

\begin{figure}
\begin{center}
\mbox{\epsfig{file=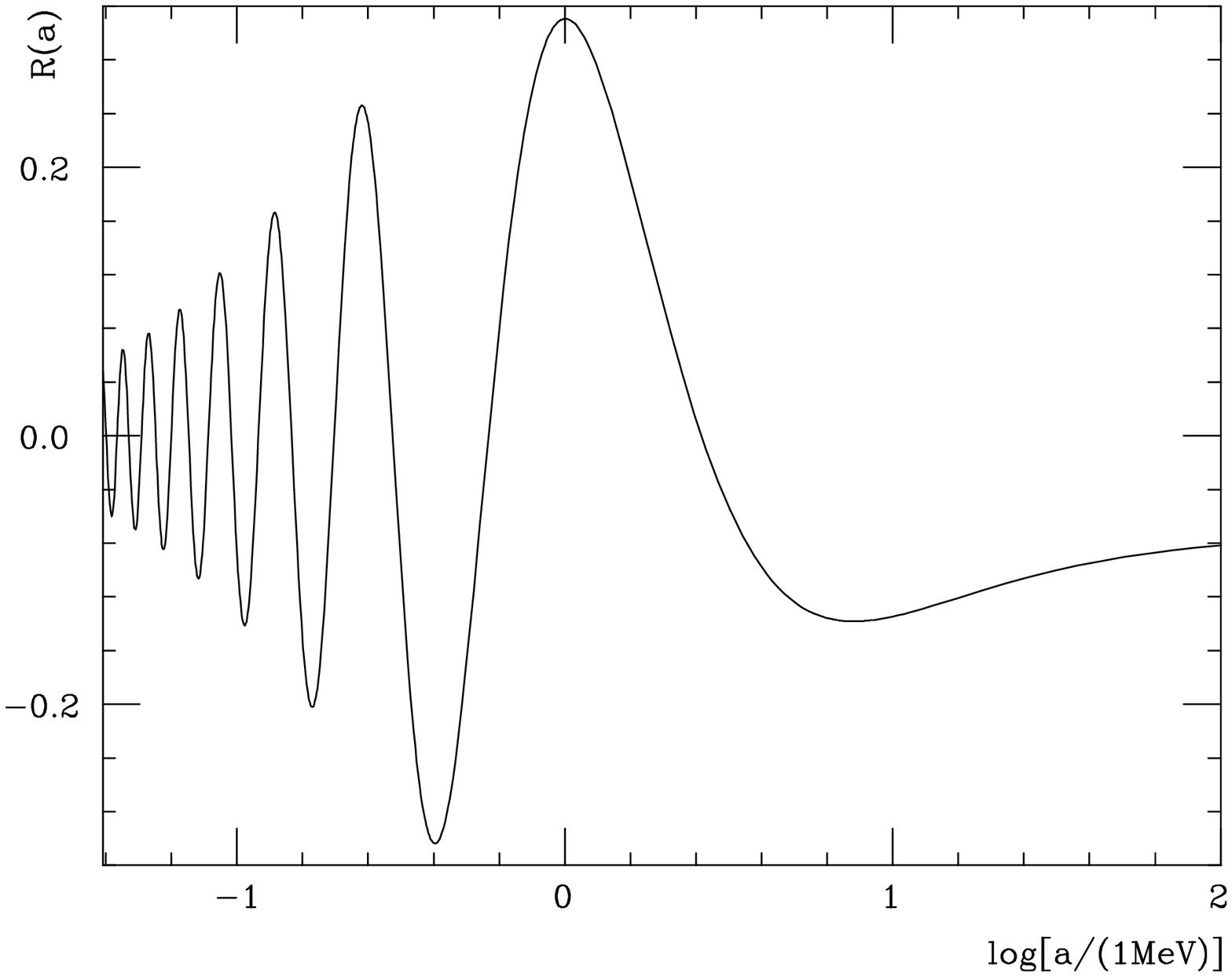,width=0.7\textwidth,angle=90}}
\end{center}
\vskip -1 cm
\caption{}
\label{FIG1}
\end{figure}

\newpage
\begin{figure}
\begin{center}
\mbox{\epsfig{file=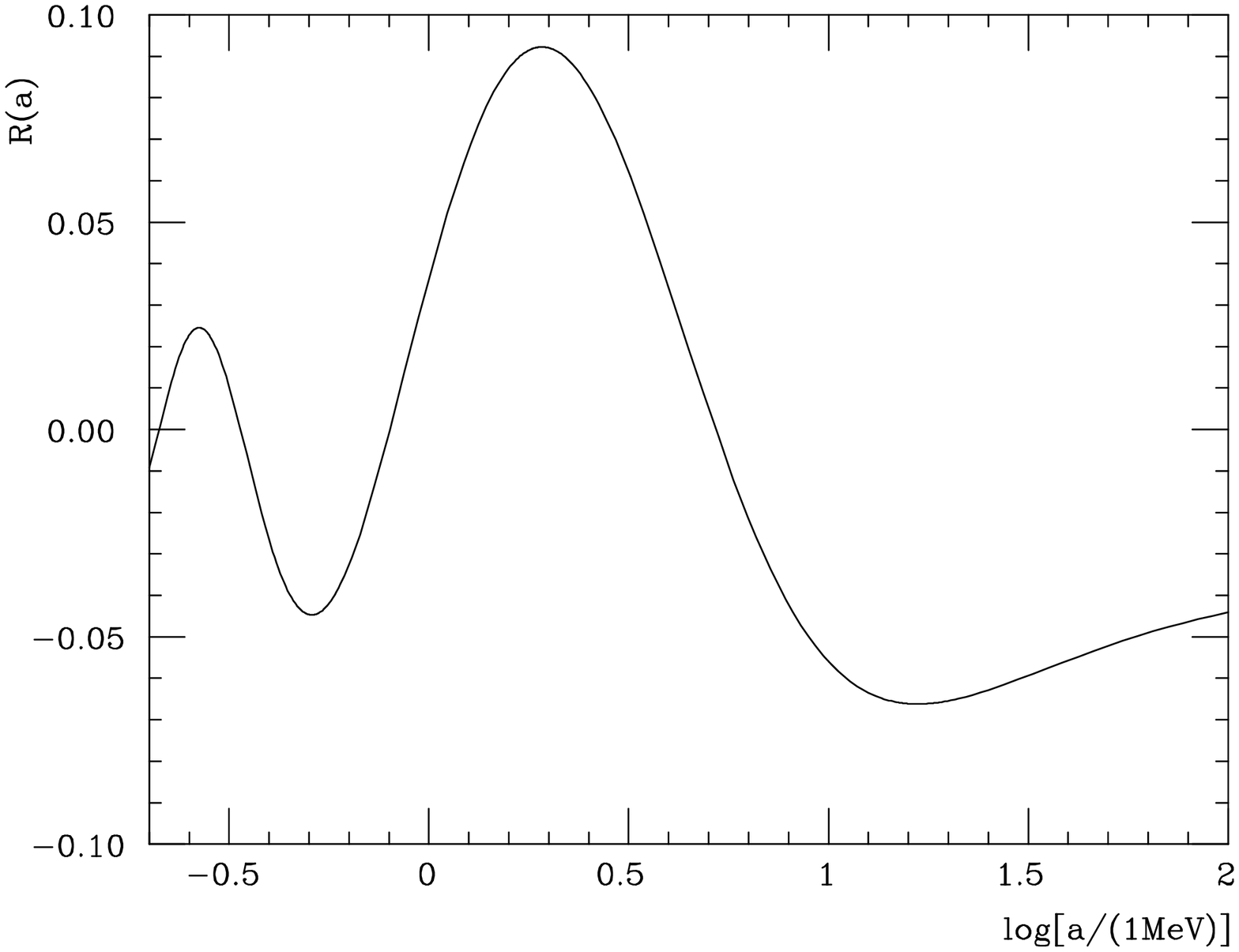,width=0.7\textwidth,angle=90}}
\end{center}
\vskip -1 cm
\caption{}
\label{FIG2}
\end{figure}

\end{document}